\documentclass[reprint,twocolumn,prl,nofootinbib,showpacs,preprintnumbers]{revtex4-1}
\usepackage{epsfig}
\usepackage{color}

\newcommand{\be}{\begin{equation}}
\newcommand{\ee}{\end{equation}}

\begin{document}
\title{High energy cosmic ray self-confinement close to extragalactic sources}
\author{Pasquale Blasi$\,{}^{1,2}$, Elena Amato$\,{}^{1}$ and Marta D'Angelo$\,{}^{2}$}
\affiliation{$^{1}$INAF-Osservatorio Astrofisico di Arcetri, Largo E. Fermi, 5 50125 Firenze, Italy}
\affiliation{$^{2}$Gran Sasso Science Institute (INFN), Viale F. Crispi 6, 60100 L'Aquila, Italy}

\date{\today}

\begin{abstract}
The ultra-high energy cosmic rays observed at the Earth are most likely accelerated in extra-galactic sources. For the typical luminosities invoked for such sources, the electric current associated to the flux of cosmic rays that leave them is large. The associated plasma instabilities create magnetic fluctuations that can efficiently scatter particles. We argue that this phenomenon forces cosmic rays to be self-confined in the source proximity for energies $E<E_{\rm cut}$, where $E_{\rm cut}\approx 10^{7} L_{44}^{2/3}$ GeV for low background magnetic fields ($B_{0}\ll nG$). For larger values of $B_{0}$, cosmic rays are confined close to their sources for energies $E<E_{\rm cut}\approx 2\times 10^{8} \lambda_{10} L_{44}^{1/4} B_{-10}^{1/2}$ GeV, where $B_{-10}$ is the field in units of $0.1$ nG, $\lambda_{10}$ is its coherence lengths in units of 10 Mpc and $L_{44}$ is the source luminosity in units of $10^{44}$ erg/s. 
\end{abstract}

\maketitle
{\it Introduction} -- The sources of ultra high energy cosmic rays (UHECRs) are yet to be found. Actually at present there is no clear consensus even on the transition energy at which cosmic rays (CRs) start being mainly of extragalactic origin \cite{Rendus}. Some lower limits on the average luminosity of UHECR sources have been presented in the literature \cite{waxman}, hinting at the need for very high luminosities. 

In the following we will not be concerned about the nature of the source. We will only assume that UHECRs are accelerated in some unspecified extragalactic sources and focus on their propagation while leaving the parent galaxy. We will refer to the latter as the ``source'' for the sake of brevity, but the actual accelerator can be anything within the parent galaxy.

Let us assume, for simplicity, that the CR sources accelerate a spectrum $q(E)\propto E^{-2}$ up to some maximum energy $E_{\rm max}$. The differential number density of CRs streaming out of such sources will read:
$$
n_{\rm CR}(E,r) = \frac{q(E)}{4\pi r^{2} c}=\frac{L_{\rm CR}}{\Lambda} \frac{E^{-2}}{4 \pi r^2c}\approx 
$$
\be
\approx 1.7\times 10^{-14} L_{44}\ E_{\rm GeV}^{-2}\ r_{\rm Mpc}^{-2}\  \rm cm^{-3} \rm GeV^{-1}\ ,  
\label{eq:nbal}
\ee
where we have taken $\Lambda=\ln(E_{\rm max}/E_{\rm min})\approx 25$ and $L_{\rm CR}=10^{44} L_{44}$ erg/s, energies are in GeV and distances in Mpc.
We assume that the source is in a region of the intergalactic medium (IGM), where the density of baryonic gas is $n_{b}=\Omega_{b}\rho_{cr}/m_{p}\approx 2.5\times 10^{-7}~\rm cm^{-3}$ (where $\rho_{cr}$ is the critical mass density of the universe and $\Omega_{b}=\rho_{b}/\rho_{cr}\simeq 0.022$ is the ratio of the baryon density $\rho_{b}$ with respect to the critical density) and we assume that there is a cosmological magnetic field with a strength  $B_{0}=10^{-13}B_{-13}$ G and a correlation scale $\lambda_{B}\sim 10 ~\rm Mpc ~\lambda_{10}$ (where $\lambda_{10}$ is in units of 10 Mpc). Hence on scales smaller than $\lambda_{10}$, the field can be considered as oriented along a given $\hat z$ direction. In such a situation, the Alfv\'en speed is $v_{A}=B_{0}/\sqrt{4\pi \Omega_{b}\rho_{cr}} \approx 44 ~B_{-13}\rm cm~s^{-1}$.

The positively charged CRs leaving their sources form an electric current that will be compensated by motions in the background plasma (return current) so as to ensure local charge and current neutrality. This situation is known to give rise to a non-resonant plasma instability that is potentially very important for CR transport. This instability was first proposed in Ref.~\cite{bell2004,bl2001} in connection with CR acceleration in supernova remnants, in which context it may provide a mechanism for strong magnetic field amplification in the shock proximity, a necessary condition to accelerate particles to $\sim PeV$ maximum energies.

The main purpose of this letter is to evaluate under which conditions this phenomenon gives rise to an instability and what the consequences are in terms of CR propagation and intergalactic magnetic field generation.
\vskip .5cm
{\it The calculations} -- The current associated with CRs streaming away from their sources in the IGM is easily written as a function of the minimum energy $E$ of particles in the current as 
\be
J_{\rm CR}=e n_{\rm CR}(>E) c=\frac{e L_{\rm CR}E^{-1}}{4\pi \Lambda r^{2}} = e\frac{q(>E)}{4\pi r^{2}},
\label{eq:current}
\ee
where in the last equality we made use of Eq. \ref{eq:nbal}.

This expression is strictly valid only if the background field is zero, or if the Larmor radius of the particles is $r_L\gg\lambda_B$, but we shall see that the above estimate for the current density turns out to hold also in the diffusive regime.

A current propagating in a plasma can give rise to instabilities of different types. Granted that the current carrying particles are well magnetised ($v_A>J_{\rm CR}/(e n_b)$, which from Eq.~\ref{eq:current} is seen to imply $B_0>2 \times 10^{-13}\ L_{44} E_{\rm GeV}^{-1} r_{\rm Mpc}^{-2}$)~G, the fastest growing instability arises when the condition 
\be
J_{\rm CR}E>\frac{ce B_0^2}{4\pi}
\label{eq:cond1}
\ee
is satisfied. This condition, which is the standard one for the development of non-resonant modes of the streaming instability \cite{bell2004}, is equivalent to the requirement that the energy density of CRs be locally larger than the energy density in the form of pre-existing magnetic field, $B_{0}^{2}/4\pi$. For $q(E)\propto E^{-2}$, this requirement becomes independent of $E$ and, using Eq.~\ref{eq:current}, can be simply formulated as:
\be
r <  r_{\rm inst} =3.7\times 10^{4} \frac{L_{44}^{1/2}}{B_{-13}}~ \rm Mpc.
\label{eq:nonres}
\ee
When Eq.~\ref{eq:nonres} is satisfied the fastest growing modes in the amplified field $\delta B$ have a wavenumber $k_{\rm max}$ that reflects the equilibrium between magnetic tension and $J_{\rm CR}\times \delta B$ force on the plasma, namely $k_{\rm max}B_{0} = \frac{4\pi}{c}J_{\rm CR}$, and their growth rate is:
\be
\gamma_{\rm max} = k_{\rm max} v_{A}=\sqrt{\frac{4\pi}{n_{b} m_{p}}} \frac{J_{\rm CR}}{c}\ ,
\label{eq:growthrate}
\ee
independent of the initial value of the local magnetic field $B_{0}$. The scale of the fastest growing modes $k_{\rm max}^{-1}$ is much smaller than the Larmor radius of the particles dominating the current (this is entailed in the condition for growth, Eq.~\ref{eq:cond1}), therefore they have no direct influence on particle scattering. This conclusion is however changed by the non-linear evolution of the modes. As long as the instability develops on small scales, it cannot affect the current, hence one could treat the two as evolving separately. A fluid element will be subject to a force that is basically $\sim J_{\rm CR}\delta B/c$: its equation of motion is $\rho (dv/dt) \simeq \frac{1}{c} J_{\rm CR} \delta B$, with $\delta B(t)=\delta B_{0} \exp\left(\gamma_{\rm max} t\right)$. As an estimate, one can write the velocity of the fluid element as $v \sim (\delta B(t) J_{\rm CR})/(c \rho \gamma_{\rm max})$, which upon integration leads to an estimate of the mean fluid displacement as $\Delta x \sim (\delta B(t) J_{\rm CR})(c \rho \gamma_{\rm max}^{2})$.
We can then estimate the saturation of the instability by requiring that the displacement equals the Larmor radius of particles in the current as calculated in the amplified magnetic field, $E/e \delta B$: when this condition is fulfilled, scattering becomes efficient and the current is destroyed. This simple criterion returns the condition:
\be
\frac{\delta B^{2}}{4\pi} \approx \frac{J_{\rm CR} E}{c e}=n_{\rm CR}(>E) E.
\label{eq:equi}
\ee
Since $n_{\rm CR}(>E)\propto E^{-1}$ in the case considered here, the saturation values of the magnetic field is independent of the energy of particles in the current driving the instability. A somewhat lower value of the saturation was inferred in \cite{riquelme}, as due to the non-linear increase of the wavelength of the fastest growing modes. Following such a prescription our saturation magnetic field would be $\sim 10$ times smaller. Eq.~\ref{eq:equi} expresses the condition of equipartition between the CR energy density and the amplified magnetic pressure, a condition that is often assumed in the literature without justification, and that here arises as a result of the physics of magnetic amplification itself. 

The field strength, as a function of the distance $r$ will read
\be
\delta B(r) = 3.7 \times 10^{-9} L_{44}^{1/2} r_{\rm Mpc}^{-1} ~\rm Gauss\ .
\label{eq:magnetic}
\ee

This rather strong magnetic field will develop over distances $r$ from the source that satisfy Eq.~\ref{eq:nonres} and under the additional condition that the growth is fast enough so as to reach saturation in a fraction of the age of the universe, $t_0$ (numerical simulations of the instability \cite{bell2004} show that saturation occurs when $\gamma_{max}\tau \sim 5-10$). This latter condition reads $\gamma_{\rm max} t_0\gtrsim 5$ and translates into:
\be
r<r_{\rm growth}=1.2 \times 10^{4}L_{44}^{1/2}E_{\rm GeV}^{-1/2}\ \rm Mpc\ .
\label{eq:growth}
\ee
If the conditions expressed by Eqs.~\ref{eq:nonres} and \ref{eq:growth} are fulfilled, then the magnetic field can be estimated as in Eq.~\ref{eq:magnetic} and since $\delta B\gg B_{0}$ and there is roughly equal power at all scales (equivalent to say that $\delta B$ in Eq. \ref{eq:equi} is independent of energy $E$), it is reasonable to assume that particle propagation can be described as diffusive, with a diffusion coefficient corresponding to Bohm diffusion in the magnetic field $\delta B$. This assumption is based on two different considerations: Bohm diffusion regime is generally found in the quasi-linear theory of wave particle interactions when $\delta B/B_0 \approx 1$ and $\delta B^2$ is roughly independent on scale (the dependence is only logarithmic: see e.g. \cite{lc83}); in the particular case of turbulence generated by Bell's instability, with $\delta B/B_{0}\gg 1$ and scale invariant power spectrum, there is additional evidence that transport is governed by Bohm diffusion (see  \cite{mlp06} and \cite{bl2001} for extensive discussion). We can then write the particle diffusion coefficient as:
\be
D(E,r)=9\times 10^{24} E_{\rm GeV}\ r_{\rm Mpc}\ L_{44}^{-1/2} \rm cm^{2}~s^{-1}.
\ee
The initial assumption of ballistic propagation of CRs escaping a source leads to conclude that particles would produce enough turbulence to make their motion diffusive. The diffusion time over a distance $r$ from the source can be estimated as
$\tau_{d}(E,r) = r^{2}/D(E,r) \approx 3.3\times 10^{16} r_{\rm Mpc}\ E_{\rm GeV}^{-1}\ L_{44}^{1/2} \rm yr$
from which follows that particles can be confined within a distance $r$ from the source for a time exceeding the age of the Universe, if their energy satisfies the condition:
\be
E\lesssim E_{\rm conf}=2.6\times 10^{6}\ r_{\rm Mpc}\ L_{44}^{1/2}\ \rm GeV\ .
\label{eq:confine}
\ee

One might argue that this conclusion contradicts the assumptions of our problem: for instance the density of particles in the diffusive regime is no longer as given in Eq.~\ref{eq:nbal}. This is certainly true, but the current that is responsible for the excitation of the magnetic fluctuations remains the same, as can easily be demonstrated: for particles with energy $E>E_{\rm conf}$ in Eq.~\ref{eq:confine}, and assuming that energy losses are negligible, quasi-stationary diffusion can be described by the equation
\be
\frac{1}{r^{2}}\frac{\partial}{\partial r}\left[ r^{2} D(E,r) \frac{\partial n}{\partial r}\right] = \frac{q(E)}{4\pi r^{2}}\delta (r),
\ee
where $q(E)$ is the injection rate of particles with energy $E$ at $r=0$. Here the advection term has been neglected since there is no bulk motion of the background plasma and the Alfv\'en speed is very small. This equation is easily solved to provide the density of CRs at distance $r$ from the source:
\be
n(E,r) \approx \frac{q}{8\pi r D(E,r)}.
\label{eq:ndiff}
\ee
Clearly, by definition of diffusive regime, the density of particles returned by Eq.~\ref{eq:ndiff} is larger than the density in the ballistic regime, Eq.~\ref{eq:nbal}. However, the current in the diffusive regime is 
\be
J_{\rm CR}^{\rm diff}= e E D(E,r) \frac{\partial n}{\partial r} = e\frac{q(>E)}{4\pi r^{2}},
\label{eq:diffcur}
\ee
which is exactly the same current that we used in the case of ballistic CR propagation (see Eq.~\ref{eq:current}). This is a very important and general result: the magnetic field in Eq.~\ref{eq:magnetic} is reached outside a CR source independent of the mode of propagation of CRs, since it is only determined by the current and not by the absolute value of the CR density. Clearly the particles that are confined within a distance $r$ around the source do not contribute to the CR current at larger distances. 
\vskip .5cm
{\it Results and implications} -- The confinement energy in Eq.~\ref{eq:confine} is somewhat ambiguous since it depends on the distance $r$. What is the highest energy at which CRs escaping a source of given luminosity are confined to the source vicinity? In order to answer this question we need to take into account all the three conditions that need to be imposed to guarantee confinement, namely Eq.~\ref{eq:nonres} (existence of fastly growing modes), Eq.~\ref{eq:growth} (growth rate faster than the expansion of the universe) and Eq.~\ref{eq:confine} (confinement). The first condition yields a limit on the distance from the source that is easy to satisfy, unless the strength of the background magnetic field is increased by several orders of magnitude, in which case however other complications arise (see discussion below). 

The other two conditions lead to the constrain
\be
E_{\rm cut} \approx 10^{7}\ {\rm GeV}\ L_{44}^{2/3} .
\label{eq:Ecut}
\ee
These particles are confined within a distance from the source:
\be
r_{\rm conf} \approx 3.8\ {\rm Mpc}\ L_{44}^{1/6} .
\label{eq:Rconf}
\ee
Within such a distance the magnetic field is as given by Eq.~\ref{eq:magnetic} and larger than $\delta B(r_{\rm conf})\approx 9.6 \times 10^{-10} L_{44}^{1/3}$ G. It is noteworthy that both the size of the confinement region and the magnetic field depend weakly upon the CR luminosity of the source, respectively as $L_{44}^{1/6}$ and $L_{44}^{1/3}$. Hence we can conclude that magnetic fields at the level of $0.1-1$nG must be present in regions of a few Mpc around the sources of UHECRs. As a consequence, the spectrum of CRs leaving these sources and eventually reaching the Earth must have a low energy cutoff at an energy $E_{cut}$. This kind of cutoff has been often postulated in the literature in order to avoid some phenomenological complications that affect models for the origin of UHECRs. For instance, a low energy cutoff is required in the dip model \cite{dip1,dip2} to describe appropriately the transition from Galactic to extragalactic CRs. This feature is usually justified by invoking some sort of magnetic horizon in the case that propagation of UHECRs is diffusive in the lower energy part of the spectrum \cite{horizon}. A similar low energy suppression of the CR flux is required by models with a mixed composition \cite{mixed}. In the calculations illustrated above, the presence of nuclei is readily accounted for, provided the current is still produced by protons (assumed to be the most abundant specie). In this case, the value of $E_{\rm cut}$ is simply shifted to $Z$ times higher energy for a nucleus of charge $Z$.

As discussed above, the condition that guarantees the existence of non-resonant modes (Eq.~\ref{eq:nonres}) is easily satisfied, unless the background magnetic field reaches $B_0\approx 9.6 \times 10^{-10}\ L_{44}^{1/3}$. However, when this happens the calculations above break down for another reason: CRs can freely stream from the source only if their Larmor radius in the pre-existing magnetic field is much larger than the assumed coherence scale of the field, namely
\be
E\gg 10^6 {\rm GeV}\ B_{-13} \lambda_{10}.
\label{eq:Larmor}
\ee
When Eq.~\ref{eq:Larmor} is not fulfilled, namely when the background field is relatively strong, then the propagation of CRs from the source becomes intrinsically one dimensional, which implies that the density of particles can be written as
\be
n_{\rm CR}(E,r) = \frac{Q(E) t}{\pi r_{L}^{2} v t} = \frac{2Q(E)}{\pi R_c(E)^2 c},
\ee
where we used the fact that the mean velocity of the particles is $v=c/2$ for a distribution of particles that is isotropic on a half plane and we assumed that particles spread in the direction perpendicular to the background field by a distance equal to $R_c(E)=\max(r_L(E), R_s)$ with $R_s$ the source size and $r_L(E)$ the Larmor radius of particles of given energy $E$. For a given source size $r_L>R_s$ as soon as $E_{\rm GeV} \gtrsim 9 \times 10^{6} B_{-10} (R_s/100 kpc)$. At energies larger than this:
\be
n_{\rm CR}(E,r)  \approx 47 L_{44}  E_{\rm GeV}^{-4} B_{-10}^{2} {\rm cm}^{-3} {\rm GeV}^{-1},
\ee
If particles with energy $>E$ are able to reach a given location, the current at that location is
\be
J_{\rm CR} \approx e \frac{c}{2} E n_{\rm CR}(E,r) = e  \frac{E Q(E)}{\pi r_{L}^{2}}\ ,
\ee
which results in non-resonant growth of the field for 
\be
E<E_{\rm inst}=3.5 \times 10^9 {\rm GeV}\ L_{44}^{1/2}\ ,
\label{eq:nonres2}
\ee
independent of $B_0$, and in a saturation magnetic field
\be
\delta B \approx 0.7\ E_{\rm GeV}^{-1}\ L_{44}^{1/2}\ B_{-10} ~ \rm Gauss.
\ee
This value of the magnetic field is apparently very large and reflects the very large density of particles at low energies in the proximity of the source, as due to the reduced dimensionality of the problem. However one should notice that the value is normalized to the density of $GeV$ particles, which only live in the immediate vicinity of the source and generate small scale fields to which high energy particles are almost insensitive. At Mpc scales, where only high energy particles can reach, the field strength is much lower as  we discuss below.

Assuming again that the diffusion coefficient is Bohm-like, one can write:
\be
D(E,r)=4.8\times 10^{16}\ E_{\rm GeV}^{2}\ L_{44}^{-1/2} B_{-10}^{-1} ~ \rm cm^{2}/s,
\ee
which leads to an estimate of the diffusion time:
$\tau_{\rm diff} = 6.2\times 10^{24}\ E_{\rm GeV}^{-2}\ L_{44}^{1/2}\ B_{-10}\ r_{\rm Mpc}^{2}~\rm yr.$
Requiring that particles reach the location at distance $r$ in a time shorter than the age of the universe, we then obtain:
\be
r_{\rm conf} \approx 0.5\  \left(\frac{E}{10^7\rm GeV}\right)\ L_{44}^{-1/4}\ B_{-10}^{-1/2}~ Mpc.
\label{eq:rconf2}
\ee
Following the usual procedure, one can calculate the growth rate of the fastest modes:
\be
\gamma_{\rm max} = \sqrt{\frac{4\pi}{\rho_{b}}}\frac{e c n_{\rm CR}(>E) }{2} = 1.9\times 10^{18} L_{44} B_{-10}^{2} E_{\rm GeV}^{-3} ~ \rm s^{-1},
\ee
and impose the condition that $\gamma_{\rm max}t_{0}>5$, which reads:
\be
E\lesssim E_{\rm growth} \approx 5.3\times 10^{11}\ {\rm GeV} L_{44}^{1/3} B_{-10}^{2/3}.
\label{eq:gmax2}
\ee

The intersection of all the conditions listed above leads to conclude that particles with energies 
\be
E<E_{\rm cut}=2.2\times 10^8 {\rm GeV}\ L_{44}^{1/4}\ B_{-10}^{1/2}\ \lambda_{10}
\ee
will be confined within a radius 
\be
r_{\rm conf}\approx 10\ {\rm Mpc}\ \lambda_{10}\ .
\ee
The amplified magnetic field at such distance is 
\be
\delta B\approx 3 \times 10^{-9} {\rm G}\ L_{44}^{1/4}\ B_{-10}^{1/2}\ \lambda_{10}^{-1}.
\ee
We emphasise again that the results illustrated both in the case of 3-d (lower $B_{0}$) or 1-d propagation (higher $B_{0}$) are only sensitive to the CR current, and hence insensitive to whether particle propagation is ballistic or diffusive. 

\vskip .5cm
{\it Discussion and Conclusions -} It is often the case that CRs affect the environment in which they propagate, through the emission and absorption of waves that couple them with the background plasma. The phenomenon of self-generation of waves is especially important close to shock fronts, where this process heavily affects the maximum energy that can be reached \cite{bell78,bell2004} and is accompanied by observational consequences \cite{blasirev,amatorev}, such as spatially thin rims of enhanced X-ray synchrotron emission (see \cite{Vink} for a review). Here we investigated these processes when CRs start their journey from extragalactic sources: the escaping CRs form an electric current to which the background plasma, at density $\Omega_{b}\rho_{cr}$, reacts, by generating a return current that in turn leads to the development of small scale instabilities. The growth of such instabilities leads to large turbulent magnetic fields and to enhanced particles' scattering. 

The details of this process depend on the strength of the pre-existing magnetic field $B_{0}$: if it is very weak (say $\lesssim 10^{-10}$ G) then, in the absence of non-linear phenomena, CRs will try to propagate in approximately straight lines. The resulting electric current leads to the development of a Bell-like instability, that modifies the propagation of particles to be diffusive: particles with energy $\lesssim 10^{7} L_{44}^{2/3}$ GeV are confined inside a distance of $\approx 3.8 L_{44}^{1/6}$ Mpc from the source for times exceeding the age of the Universe, thereby introducing a low-energy cutoff at such energy in the spectrum of CRs reaching the Earth. Since the confinement distance is weakly dependent on the source luminosity, we conclude that a region with $\sim nG$ fields should be present around any sufficiently powerful CR source. If larger background magnetic fields are present around the source, the gyration radius of the particles can be smaller than the coherence scale of the field, and in this case CR propagation develops in basically one spatial dimension. For a coherence scale of 10 Mpc, CRs are confined in the source proximity for energies $E\lesssim 2\times 10^{8} L_{44}^{1/4} B_{-10}^{1/2}\ \lambda_{10}$ GeV. 

It is currently not known whether the confinement phenomenon occurs in one or the other regime since only limits exist on cosmological magnetic fields: upper limits can be obtained from Faraday rotation measures \cite{burles} but these limits are rather weak ($\lesssim nG$) and model dependent. A lower limit can be found from gamma ray observations of distant TeV sources \cite{nero1,nero2} and these limits are typically at the level $B_{0}\gtrsim 10^{-17}$ G. Numerical simulations of large scale structure formation in the presence of background magnetic fields typically find $\sim 10^{-13}$ Gauss magnetic fields in voids \cite{dolag} (although see \cite{miniati} for different conclusions). 

The physical prescription adopted here leads to estimating the strength of the self-generated magnetic field $\delta B$ at the level of equipartition with the energy in the form of escaping cosmic rays, independent of the value of the pre-existing field, $B_{0}$. A weak dependence on $B_{0}$ was instead found for the saturation level in \cite{riquelme}, which in our case would lead to $\delta B$ about $\sim 10$ times smaller for small values of $B_{0}$, thereby reducing the energy below which CRs are confined in the source proximity. Understanding the dynamics of the magnetic field amplification and saturation is clearly very important. One could test the amplification mechanism in the case of supernova remnant shocks: in this case the saturation criterion used here translates to $\delta B\approx \sqrt{4 \pi w_{\rm CR} v_S/c}$, with $v_S$ the velocity of the supernova blast wave and $w_{\rm CR}$ is the energy density in accelerated particles. Applying this criterion, we obtain an estimate of the magnetic field which is in good agreement with that measured in young galactic SNRs \cite{Vink}. On the other hand, due to the relatively small value of $\delta B/B_{0}$, the saturation provided by \cite{riquelme} would return a value of $\delta B$ only a factor $\sim 2$ smaller, too small a difference to discriminate between the two estimates. The testing is then left to numerical experiments studying the propagation of a current of energetic particles in a low density, low magnetic field plasma: hybrid simulations with this aim are currently ongoing.

The phenomenon of CR confinement illustrated here has profound implications for the description of the transition region between Galactic and extra-galactic CRs \cite{dip1,dip2,mixed}. It is rather tantalising that the cutoff obtained here as due to self-trapping is in the same range of values that have previously been invoked in the literature based upon phenomenological considerations. 

\vskip .5cm
{\it Acknowledgments} -- We are very grateful to D. Caprioli, P. Serpico and L. Sironi for useful comments. This work was partially funded through grant PRIN-INAF 2013.

\end{document}